\documentclass[conference]{IEEEtran}

\usepackage{amsmath}
\usepackage{amsthm}
\usepackage{amssymb}
\usepackage{graphicx}
\usepackage{subcaption}
\usepackage{url}
\usepackage{epstopdf}
\usepackage{placeins}
\usepackage{epsfig}
\usepackage{bbm}
\usepackage{amsfonts}
\usepackage{balance}
\usepackage{graphicx}
\usepackage{xfrac}
\usepackage{mathrsfs}
\usepackage{algorithm} 
\usepackage{algorithmic}
\usepackage{footnote}
\usepackage[table]{xcolor}
\usepackage{mathtools,amssymb,lipsum, nccmath}
\usepackage{cuted}

\makeatother

\newtheorem{theorem}{Theorem}
\newtheorem{definition}{Definition}
\newtheorem{lemma}{Lemma}
\newtheorem{example}{Example}
\newtheorem{remark}{Remark}
\newtheorem{corollary}{Corollary}
\newcommand{\mytilde}{\raise.17ex\hbox{$\scriptstyle\mathtt{‌​\sim}$}}
\usepackage{multirow}

\begin{document}
	\title{Minrank of Embedded Index Coding Problems and its  Relation to Connectedness of a Bipartite Graph}
	
	\author{%
		\IEEEauthorblockN{Anjana A Mahesh, and B. Sundar Rajan\\}
		\IEEEauthorblockA{ Department of Electrical Communication Engineering, Indian Institute of Science, Bengaluru 560012, KA, India \\
			E-mail: \{anjanamahesh,bsrajan\}@iisc.ac.in}
	}

	%	\markboth{IEEE Transactions on Vehicular Technology,~Vol.~XX, No.~XX, XXX~2019}
	{}

	\maketitle
	
	\begin{abstract} 
		This paper deals  with embedded index coding problem (EICP), introduced by A. Porter and M. Wootters, which is a decentralized communication problem among users with side information. An alternate definition of the parameter minrank of an EICP,  which has reduced computational complexity compared to the existing definition, is presented. A  graphical representation for an EICP  is given using directed bipartite graphs, called bipartite problem graph, and the side information alone is represented using an undirected bipartite graph called the side information bipartite graph. Inspired by  the well-studied single unicast index coding problem (SUICP), graphical structures, similar to cycles and cliques in the side information graph of an SUICP, are identified in the side information bipartite graph of a single unicast embedded index coding problem (SUEICP). Transmission schemes based on these graphical structures, called tree cover scheme and bi-clique cover scheme are also presented for an SUEICP. Also, a relation between connectedness of the side information bipartite graph and the number of transmissions required in a scalar linear solution of an EICP is established. 
	\end{abstract}
	
	
	\IEEEpeerreviewmaketitle
	
	%%%%%%%%%%%%%%%%%%%%%%%%%%%%%%%%%%%%%%%%%%%%%%%%%%%%%%%%%%%%%%%%%%%%%%%%%
	\section{INTRODUCTION}
	\label{sec:Intro}
	
	\IEEEPARstart{M}otivated by applications in device to device multicast and distributed computing, the problem of embedded index coding was introduced by Potter and Wootters in \cite{PW}, which is a decentralized version of another well-studied informed source coding problem called the index coding problem. The index coding problem (ICP) was introduced by Birk and Kol  \cite{BiK} as an instance of informed source coding problem over a wireless broadcast channel  and has been studied extensively ever since. In an ICP, a central server which possesses a library of messages, $\mathcal{X} = \{x_1,x_2,\cdots, x_m\}$, tries to satisfy the message requests of a set of user nodes, $\mathcal{U} =\{u_1,u_2,\cdots,u_n\}$, each of which possesses some of the messages with the server a priori as side-information. It was shown in \cite{BiK} that if the server sends coded transmissions, where the messages are coded among themselves, the number of transmissions required to satisfy the message requests of all the nodes can be less than the number of distinct messages demanded. 
	
	The embedded index coding problem (EICP) differs from ICP in the fact that there is no centralized server present in the system. The user nodes themselves act as both senders and receivers and hence the only messages which can be requested by a user are those that are present with at least one other user. Instances of the EICP arise in the shuffle phase of MapReduce-like distributed computation systems \cite{SMYA}, the delivery phase of a device to device (D2D) coded caching system \cite{JCM}, the cooperative data exchange problem \cite{RSS}, which is a special case of EICP where each user requests all the messages it doesn't have, etc. The solution to an EICP is called an embedded index code. 
	
	Several graphical representations have been given for an ICP over the years, starting with modeling a symmetric instance of the ICP with $n$ messages, $n$ users with each user $u_i$ demanding a single message $x_i$ and the restriction that if the user $u_i$ demanding the message $x_i$ knows $x_j$, then the user $u_j$ demanding the message $x_j$ knows $x_i$, using an undirected graph on $n$ vertices, called side information graph, in \cite{BBJK1}. This was later extended in \cite{BBJK} to include asymmetric demands by employing directed graphs. Even this modified representation had the restriction that the number of messages and the number of users are equal and that each user demands a distinct message. This restriction was done away with in the hypergraph representation given by Dau et al in \cite{DSC} which now was applicable for any general ICP. Tehrani et al in \cite{TDN} gave an alternate representation for a general ICP using bipartite graphs. These graphical representations were used to develop transmission schemes as well as study the connection between graph parameters and the minimum number of transmissions required for a given ICP. 
	
	For a single unicast ICP (SUICP) \cite{OngHo}, which is an instance of the ICP with the number of users equal to the number of messages and where each user demands a distinct message, the graphical representation, which was given in \cite{BBJK} and called the  side-information graph, was used to derive a graph functional called the \emph{minrank} which characterized the minimum number of transmissions in a scalar linear solution to the given instance of SUICP. A scalar linear solution to an ICP or a scalar linear index code is a set of coded transmissions where each transmission is a linear combination of one generation of the messages in the system and such a solution is called optimal if the number of transmissions in it is the minimum possible. It was shown in \cite{RP} that computing minrank of a general graph is an NP-hard problem. Dau et al in \cite{DSC} extended the notion of minrank of side information graph to the minrank of directed hypergraph which characterized the length of an optimal scalar linear solution of an instance of a general ICP.
	
	Minrank of an EICP was defined in \cite{MKR} using a matrix representation of the EICP, called the side information matrix, and was shown to be equal to the length of an optimal scalar linear embedded index code. In this paper, we give an alternate definition of minrank of an EICP consistent with the definition in \cite{MKR} but along the lines of the definition of minrank of a general ICP given in \cite{DSC}. A comparison between the  definition in \cite{MKR} and that in this paper is given in section \ref{subsec:CompMR} where we explain that the computational complexity of calculating the minrank for an EICP by following the approach in this paper is much less than that required by the approach in \cite{MKR}.  
	
	A graphical representation for an EICP was given in \cite{PW} using a directed graph called problem graph. However, this representation is merely an alternate representation of the general ICP, where the same message can be requested by multiple users, as explained in section \ref{sec:Graph}. In this paper, we represent an instance of an EICP using directed bipartite graphs, as in \cite{TDN}, which we call the bipartite problem graph. We explain how the representation of an EICP using the bipartite problem graph is more intuitive and insightful than that using the problem graph in \cite{PW} in section \ref{subsec:CompGraph}. Given the side information at the user nodes, there are multiple EICPs with this side information corresponding to different demand vectors. The side information at the users is  represented using an undirected bipartite graph which we call the side information bipartite graph so as not to confuse with the side information graph of an ICP. 
	
	For an SUICP \cite{OngHo}, graphical structures like cycles and cliques, if present in the directed side information graph, were identified to give advantage in reducing the number of transmissions required and hence simple transmission schemes which makes use of these graphical structures \cite{BiK,CASL,BKL} were also presented. We identify equivalent graphical structures in the side information bipartite graph of single unicast EICPs which will help in reducing the number of transmissions required and also propose transmission schemes utilizing these graph structures.
	
	We also establish a relation between the connectedness of the side-information bipartite graph and the minrank of the EICPs which have the same side information as that represented by the side information bipartite graph. This has been motivated by the application of embedded index coding in the V2V phase of collaborative message dissemination protocol of Vehicular Adhoc Networks (VANETs) \cite{VANET,VICP}. VANETs are formed by vehicles that are within hundred meters of each other moving at low speeds in the same direction. The side information at these vehicles is obtained from an earlier R2V communication phase where a road-side unit  transmits to vehicles in its range. Since the vehicles are close-by and moving very slowly, the possibility of them receiving a lot of packets in common is quite high and hence the side information bipartite graph will be heavily connected.

	The following is a list of our technical contributions in this paper. 
	\begin{itemize}
		\item We give an alternate definition for minrank of an EICP which is computationally more efficient and prove that the length of an optimal linear solution to an EICP is equal to the minrank. 
		\item A graphical representation of an EICP is given using directed bipartite graphs with one partite set labeled by the user indices and the other by the messages, which we call as the bipartite problem graph. The side information alone is represented using an undirected bipartite graph, called the side information bipartite graph.  
		\item Two transmissions schemes, called tree cover scheme and bi-clique cover scheme, are introduced for an SUEICP, which make use of certain graphical structures called regular trees and bi-cliques. 
		\item A theorem which establishes a relation between the connectedness of the side information bipartite graph and the minrank of the EICPs which have the same side information as that represented by the side information bipartite graph is presented.
		
	\end{itemize}

	The rest of this paper is organized as follows. The embedded index coding problem is formally set up in section \ref{sec:Problem}. This is followed by a definition of minrank of an EICP and a relation between minrank of an EICP and the length of an optimal linear solution to the EICP in section \ref{sec:Minrank}. A graphical representation for an EICP is given, and graph structures called regular trees and bi-cliques identified in the side information bipartite graph for an SUEICP and transmission schemes utilizing these graph structures are presented in section \ref{sec:Graph}. A theorem which establishes a relation between the connectedness of a side information bipartite graph and the minrank of the EICPs which have their side information as that represented by the side information bipartite graph is given in section \ref{sec:ConMinrank}.  Finally the paper is concluded in section \ref{sec:Conc} by summarizing the contributions in the paper and identifying problems for future research.
	
	\emph{Preliminaries} :  For an $n$-set $X=\{x_1,x_2,...x_n\}$ and a subset  $A=(a_1,a_2,...a_m)$ of $\{1,2,\cdots,n\}$, the restriction of $X$ to the set $A$ is the subset $\{x_{a_1},x_{a_2},...x_{a_m}\}$ of $X$. Similarly for a vector of length $n$, $\mathbf{x}=(x_1,x_2,...x_n)$, its restriction to a subset $A=(a_1,a_2,...a_m)$ of $\{1,2,\cdots,n\}$ is the vector $(x_{a_1},x_{a_2},...x_{a_m})$ of length $m \leq n$. The support of a vector $\mathbf{x}$ of length $n$ is the set of its non-zero indices, i.e., $supp(\mathbf{x}) = \{i \in \{1,2,\cdots,n\} \text{ s.t } x_i \neq 0 \}$.  The unit vector in $\mathbb{F}^n_q$ with a $1$ in the $i^{\text{th}}$ position and zero elsewhere is denoted as $\mathbf{e_i} = (\underbrace{0,\cdots,0}_{i-1 \text{ zeros}},1,\underbrace{0,\cdots,0}_{n-i \text{ zeros}})$. An indicator function $\mathcal{I}(x)$ takes the value $1$ if $x$ evaluates to true and $0$ otherwise. \\

	\emph{Notations}: For a prime power $q$, $\mathbb{F}_{q}$ denotes the finite field with $q$ elements. For a positive integer $n$, $[n]$ denotes the set $\{1,2,\cdots,n\}$. The set of positive integers is denoted by $\mathbb{Z}^+$. A $t$-subset of $[n]$ is a subset of $[n]$ of size $t$.  For an $n$-set $X$ and a subset  $A$ of $[n]$, $X_A$ denotes the restriction of $X$ to the set $A$. Similarly for an $n$-length vector $\mathbf{x}$, $\mathbf{x}_A$ denotes the restriction of $\mathbf{x}$ to the set $A$. For an ordered set $A$ with $n$ elements, $A(j)$ is the $j^{\text{th}}$ element of $A$, for $j \in [n]$, whereas, for an $m \times n$ matrix $A$, $A(j)$ is the $j^{\text{th}}$ row of $A$, for $j \in [m]$. For an $n$-length vector $\mathbf{x}$ and a set $S \subset [n]$, $\mathbf{x} \triangleleft S$ indicates that the vector $\mathbf{x}$ has its support in $S$. The transpose of a vector $\mathbf{v}$ is denoted as $\mathbf{v}^T$ and that of a matrix $A$ is denoted as $A^T$. 
	%For two positive integers $m$ and $n$, the notation $<m>_n$ is used to mean the value of $m$ modulo $n$. 
	% The symbol $\oplus$ denotes the XOR of its operands. 
	% Also, $\binom{n}{k}= \frac{n!}{k!(n-k)!}$ and $\binom{n}{k}=0$, when $n < 1$ or $n < k$.  
	%%%%%%%%%%%%%%%%%%%%%%%%%%%%%%%%%%%%%%%%%%%%%%%%%%%%%%%%%%%%%%%%%%%%%%%%%%%%%%%%%%%%%%%%%%%%%%%%%%%%
	
	\section{Problem Setup}
	\label{sec:Problem}
	
	Consider an embedded index coding problem with $N$ users, $\mathcal{U} = \{u_1,u_2, \cdots, u_N\}$ and a set of $M$ messages $\mathcal{X} = \{x_1,x_2,\cdots x_M\}, \ x_i \in \mathbb{F}_q$, with  user $u_i$ demanding a subset of messages $\mathcal{W}_i \subset \mathcal{X}$ and possessing a non-intersecting subset of messages, indexed by an ordered set $\mathcal{K}_i$, as side-information. The goal of the EICP is to satisfy the message requests of all the users with minimum number of transmissions by the users themselves. There is no central server which possesses all the messages in $\mathcal{X}$ and hence the side-information possessed by the users is such that $\bigcup\limits_{i=1}^{K} \mathcal{X}_{\mathcal{K}_i} = \mathcal{X}$, i.e., every message is present with at least one user. No user possesses all the messages, i.e.,  $\mathcal{X}_{\mathcal{K}_i} \subsetneq \mathcal{X}, \ \forall i$, since then that user can act as a central server and the EICP reduces to the centralized ICP and any solution of the centralized ICP can be transmitted by this user which possesses all the messages. Further, it is also assumed that no message is available at all users as then that message won't be demanded by any user and can as well be removed from the system. 
	
	Since a user demanding $k$ messages can be split into $k$ users each demanding a single message and all the $k$ users having the same side information as the original user, in the rest of this paper, we consider that each user demands a single message in an EICP. Let the message demanded by a user $u_i$ be denoted as $x_{d_{i}}$, where $d_{i} \in [M]$ and let the vector $\mathbf{d}$ denote the vector of indices of messages demanded by all the $N$ users, i.e., $\mathbf{d} = (d_1,d_2,\cdots, d_N)$. Further, let the side information possessed by all the users be denoted by the set $\mathcal{K} = \{\mathcal{K}_1,\mathcal{K}_2, \cdots, \mathcal{K}_N\}$. An EICP with $N$ users, $M$ messages, $M \leq N$, side information set $\mathcal{K}$ and demand vector $\mathbf{d}$ is denoted as $\mathcal{E}(N,M,\mathcal{K},\mathbf{d})$. 
	
	A solution to an EICP, called an embedded index code, is a set of transmissions made by all or a subset of the user nodes such that the demands of all the users are satisfied. An embedded index code is called linear if all the transmissions involved are linear combinations of the messages. An optimal embedded index code is one with minimum number of transmissions. An embedded index code over $\mathbb{F}_q$ for an EICP $\mathcal{E}(N,M,\mathcal{K},\mathbf{d})$ is defined as follows. 
	
	\begin{definition}
		For an EICP $\mathcal{E}(N,M,\mathcal{K},\mathbf{d})$ over $\mathbb{F}_q$, an embedded index code consists of 
		\begin{enumerate}
			\item a set of encoding functions $\{\mathcal{C}_i \}_{i\in T_u}$, where $T_u \subseteq [N]$ is the index set of transmitting users, such that at a user $u_i \in \mathcal{U}_{T_u}$, the encoding function $C_i: \mathbb{F}_q^{|\mathcal{K}_i|} \rightarrow \mathbb{F}_q^{l_i}$ encodes the messages in its side information into $l_i$ coded messages over $\mathbb{F}_q$, and
			\item a set of $N$ decoding functions, $\{\mathcal{D}_i \}_{i\in [N]}$, one at each of the $N$ users, such that at a user $u_i$, the decoding function $\mathcal{D}_i : \mathbb{F}_q^{\sum_{j \in T_u \setminus \{i\}}l_j} \times \mathbb{F}_q^{|\mathcal{K}_i|} \rightarrow \mathbb{F}_q$ such that $\mathcal{D}_i \Big(\{\mathcal{C}_j \}_{j\in T_u \setminus \{i\}}, \mathcal{X}_{\mathcal{K}_i}\Big) = x_{d_i}$.
		\end{enumerate}
	\end{definition}  
	
	The length of the embedded index code defined above is $l = \sum\limits_{j \in T_u}l_j$. 
	
	\begin{definition}
		An embedded index code is said to be linear if all the encoding functions involved are linear transformations over $\mathbb{F}_q$. In such an embedded index code, each of the encoding functions in $\{\mathcal{C}_i \}_{i\in T_u}$, can be represented using a $|\mathcal{K}_i| \times l_i$ matrix $L_{i}$ with entries from $\mathbb{F}_q$. 
	\end{definition}
	
	\begin{example}
		\label{ex:EICP1}
		Consider an EICP with $N=4$ users and $M=4$ messages, $\mathcal{X} = \{x_1,x_2,x_3,x_4\}$. Let the side information at the users be $\mathcal{K}_1 = \{1\}, \ \mathcal{K}_2 = \{1,2,3\}, \ \mathcal{K}_3 = \{2,4\}, \ \mathcal{K}_4 = \{4\}$ and the demand vector be $\mathbf{d} = (2,4,1,3)$. For this EICP, an optimal linear solution corresponding to the transmitting users $\mathcal{U}_{T_u} = \{u_2,u_3\}$ is given by $\mathcal{C}_2 = \begin{bmatrix}
			1 & 1 & 0  \\ 	
			0 & 0 & 1 
		\end{bmatrix}^T, \ \mathcal{C}_3 = \begin{bmatrix}
			0 & 1
		\end{bmatrix}^T$. For the message vector $\mathbf{x} = (x_1,x_2,x_3,x_4)$, the transmissions corresponding to this index code of length $l =3$ are 
		$\mathbf{x}_{\mathcal{K}_2}\mathcal{C}_2 = \begin{bmatrix}
			x_1+x_2\\x_3
		\end{bmatrix}$ and  $\mathbf{x}_{\mathcal{K}_3}\mathcal{C}_3 = x_4 $. It can be verified that each of the user nodes can decode their demanded messages from the transmissions using their side information. 
		
	\end{example}
	In \cite{MKR}, a matrix representation of an EICP was given and a parameter called minrank was derived from this matrix representation which characterized the length of an optimal scalar linear embedded index code. In the following section, an alternate definition of minrank is proposed and a proof that the minrank characterizes the length of an optimal embedded index code is given along the lines of the proof in \cite{DSC}. Also, in subsection \ref{subsec:CompMR}, we explain why the definition of minrank in this paper is better than the existing definition. 
	%%%%%%%%%%%%%%%%%%%%%%%%%%%%%%%%%%%%%%%%%%%%%%%%%%%%%%%%%%%%%%%%%%%%%%%%%%%%%%%%%%%%%%%%%%%%%%%%%%%%%
	
	\section{Minrank of an EICP}
	\label{sec:Minrank}
	As explained in the previous section, any linear solution to an EICP can be represented using a set of $|T_u|$ matrices, $\{L_i\}_{i \in T_u}$ over $\mathbb{F}_q$, where $T_u$ is the index set of transmitting users. Let $L_i^{\prime}$ be an $M \times l_i$ matrix obtained from the $|\mathcal{K}_i| \times l_i$ matrix $L_i$ as: 
	$$L_i^{\prime}(j) = \begin{cases}
		L_i(k) \text{ if } j \in  \mathcal{K}_i \text{ and }  j = \mathcal{K}_i(k). \\
		\mathbf{0} \text{ if } j \notin  \mathcal{K}_i
	\end{cases},$$
	where, $\mathbf{0}$ is a zero-row of length $l_i$. Further, let $L = [L_{i_1}^{\prime} \ L_{i_2}^{\prime} \ \cdots L_{i_{|T_u|}}^{\prime}]$, where $T_u = \{{i_1},{i_2},\cdots, i_{|T_u|}\}$, be the $M \times l$ matrix obtained by the column concatenation of the matrices in $\{L_i^{\prime}\}_{i \in T_u}$. Now, $L$ represents the embedded index code with the columns in $L_{i_j}^{\prime}$ being transmitted by the user $u_{i_j} \in \mathcal{U}_{T_u}$. Each of the columns in $L$ is linearly independent of each other as no new information will be conveyed by a transmission which can be obtained as a linear combination of other transmissions. Hence, the rank of the matrix $L$ over $\mathbb{F}_q$ is equal to the number of columns in $L$ which is equal to $l$. Therefore, an optimal linear solution will correspond to a matrix $L$ with minimum rank over $\mathbb{F}_q$ among all the matrices representing linear solutions of the given EICP. This is the idea of minrank of an EICP. 
	
	Consider an ICP with a single sender which possesses a set of $m$ messages $X = \{x_1,x_2,\cdots,x_m\}$, $x_i \in \mathbb{F}_q$,  and a set of $n$ receivers $R_1, R_2, \cdots, R_n$. A receiver $R_i$ has a subset of messages, $\mathcal{K}_i \subset X$, as side information and requests a message $x_{d_i}$. The ICP with $m$ messages, $n$ receivers, side information set $\mathcal{K} = \{\mathcal{K}_1, \mathcal{K}_2,\cdots, \mathcal{K}_n\}$ and demand vector $\mathbf{d} = (d_1,d_2,\cdots,d_n)$ is denoted as $\mathcal{H}(n,m,\mathcal{K},\mathbf{d})$. For this ICP, the definition of minrank from \cite{DSC} is given below.
	
	\begin{definition}[Minrank of an ICP \cite{DSC}]
		Suppose $\mathcal{H} = \mathcal{H}(n,m,\mathcal{K},\mathbf{d})$ corresponds to an instance of the ICP. Then the minrank of $\mathcal{H}$ over $\mathbb{F}_q$  is defined as 
		$\kappa_q(\mathcal{H}) \triangleq ~ \min\big\{\text{rank}_{\mathbb{F}_q}\big(\{\mathbf{e_{d_i}} + \mathbf{v_i}\}_{i \in [n]}\big) : \mathbf{v_i} \in \mathbb{F}_{q}^m,  \mathbf{v_i} \triangleleft \mathcal{K}_i$.	
	\end{definition}
	
	Having defined minrank of an ICP, the authors in \cite{DSC} proved that the smallest possible length of a linear index code for the ICP $ \mathcal{H}(n,m,\mathcal{K},\mathbf{d})$ over $\mathbb{F}_q$ is equal to the minrank $\kappa_q(\mathcal{H})$. Similarly, we give the following definition for minrank of an EICP and prove that it is indeed equal to the length of an optimal linear embedded index code. 
	
	\begin{definition}[Minrank of an EICP]
		For an embedded index coding problem $\mathcal{E}(N,M,\mathcal{K},\mathbf{d})$, the minrank of $\mathcal{E}$ over $\mathbb{F}_q$ is defined as 	
		\begin{equation}
			\label{eq:minrank}
			\begin{split}
				\kappa_q(\mathcal{E}) & \triangleq ~ \min\big\{\text{rank}_{\mathbb{F}_q}\big(\{\mathbf{e_{d_i}} + \mathbf{v_i}\}_{i \in [N]}\big) : \mathbf{v_i} \in \mathbb{F}_{q}^M,  \mathbf{v_i} \triangleleft \mathcal{K}_i \\  
				& \text{ s.t. } \forall i \in [N], \ \exists j \in [N],  \ j \neq i, \text{ s.t } (\mathbf{e_{d_i}} + \mathbf{v_i}) \triangleleft \mathcal{K}_j \big\}
			\end{split}
		\end{equation}	
	\end{definition}
	
	\begin{remark}
		The definition of minrank is similar to that in \cite{DSC} except for the extra condition that for each of the vectors in the set $\{\mathbf{e_{d_i}} + \mathbf{v_i}\}_{i \in [N]}$, there should be a user who has all the messages in the support set of that vector in its side information set. So, the minimization of rank is only over those sets $\{\mathbf{e_{d_i}} + \mathbf{v_i}\}_{i \in [N]}$ where for each of the vectors $\mathbf{e_{d_i}} + \mathbf{v_i}$ in the set, there exists at least one user which could transmit the corresponding coded message.
	\end{remark}
	
	\begin{theorem}
		For a given EICP $\mathcal{E}(K,N,\mathcal{K},\mathbf{d})$, the length of an optimal linear embedded index code is equal to the minrank $\kappa_q(\mathcal{E})$.
		\begin{proof}
			The proof follows along the lines of the proof in \cite{DSC}. Consider the message vector $\mathbf{x} = (x_1,x_2,\cdots, x_M)$. From a transmission of the form $T_i = \mathbf{x}(\mathbf{e_{d_i}} + \mathbf{v_i})^T$, $\mathbf{v_i} \triangleleft \mathcal{K}_i$, user $u_i$ can decode its demanded message $x_{d_i}$ as  $T_i - \mathbf{x}\mathbf{v_i}^T$. Thus , if there are $N$ transmissions $\{T_i\}_{i=1}^N$, all the users in $\mathcal{U}$ can decode their demanded messages. It is, in fact, sufficient to have $\text{rank}_{\mathbb{F}_q}\big(\{\mathbf{e_{d_i}} + \mathbf{v_i}\}_{i \in [N]}\big)$ transmissions. However, for each $i \in [N]$, there should be a user $u_j, j\neq i$ which can transmit $T_i$, i.e., $u_j$ should have all the messages involved in $T_i$ in its side information. If we only consider sets $\{\mathbf{e_{d_i}} + \mathbf{v_i}\}_{i \in [N]}$ where each of the elements $(\mathbf{e_{d_i}} + \mathbf{v_i})$ satisfies the condition that there exists some user which could transmit $T_i = \mathbf{x}(\mathbf{e_{d_i}} + \mathbf{v_i})^T$, and perform minimization of the rank over $\mathbb{F}_q$ of these sets, the minimum rank obtained will be equal to $\kappa_q(\mathcal{E})$. Since any linear embedded index code consists of a set of transmissions of the form $T_i$, the optimal length of a linear embedded index code for the given EICP $\mathcal{E}$ is equal to $\kappa_q(\mathcal{E})$. 
		\end{proof}
	\end{theorem}
	
	\subsection{Comparison with the Minrank Definition in \cite{MKR}}
	\label{subsec:CompMR}
	To compare the definitions of minrank in \cite{MKR} and in this paper, we reproduce some of the relevant definitions in \cite{MKR} for quick reference. 
	For an EICP  $\mathcal{E}(N,M,\mathcal{K},\mathbf{d})$,  the side information matrix was defined in \cite{MKR} as $M \times \sum_{i \in [N]}|\mathcal{K}_i|$ matrix $A(\mathcal{E})$ such that the  first $|\mathcal{K}_1|$ columns represented the side information at user $1$, the next $|\mathcal{K}_2|$ columns represented the side information at user $2$ and so on, as given by the formal definition below. 
	\begin{definition}[Side information matrix \cite{MKR}]
		A side-information matrix for the EICP  $\mathcal{E}(N,M,\mathcal{K},\mathbf{d})$ over $\mathbb{F}_q$ is the $M \times \sum_{i \in [M]}|\mathcal{K}_i|$ matrix $A(\mathcal{E})$ such that 
		$$ A(\mathcal{E}) = \left[\underbrace{A^{(1)}}_{|{\mathcal{K}_1| \text{ columns}}}|\underbrace{A^{(2)}}_{|{\mathcal{K}_2| \text{ columns}}}|\cdots|\underbrace{A^{(n)}}_{|{\mathcal{K}_n| \text{ columns}}} \right],$$
		where, $A^{(k)}, \ k \in [n]$ is an $m \times |\mathcal{K}_k|$ matrix $\{a_{i,j}^{(k)}\}$, $i \in [m], \ j \in [|\mathcal{K}_k|]$ with $$a_{i,j}^{(k)} = \begin{cases}
			& x \ \text{if} \ x_i \in \mathcal{K}_k \\
			& 0 \ \text{otherwise}.
		\end{cases}$$
	\end{definition}  
	A matrix $\hat{A}$ is said to be a completion of $A(\mathcal{E})$, denoted as $\hat{A} \sim A(\mathcal{E})$, if $\hat{A}$ can be obtained from $A(\mathcal{E})$ by replacing each of the $x$s in $A(\mathcal{E})$ by elements from $\mathbb{F}_q$. 
	The definition of minrank of an EICP in \cite{MKR} is as given below
	\begin{definition}[Minrank \cite{MKR}]
		The min-rank of an EICP $\mathcal{E}(N,M,\mathcal{K},\mathbf{d})$ is defined as $$\kappa_q(\mathcal{E})  = \min_{\hat{B} \sim B(\mathscr{I}(\mathcal{E}))}\min_{\hat{A} \in \hat{\mathcal{A}}(\hat{B})}\{rank_{\mathbb{F}_q}(\hat{A})\},$$ where,
		$\hat{\mathcal{A}}(\hat{B}) \triangleq \left\{ \hat{A} \sim A(\mathcal{E}) : rank_{\mathbb{F}_q}(\hat{A}) = rank_{\mathbb{F}_q}(\hat{A}|\hat{B})\right\}$ and $(\hat{A}|\hat{B})$ is the matrix obtained by the concatenation of the columns of $\hat{A}$ to the columns of $\hat{B}$.
	\end{definition}
	In the definition above, $\mathscr{I}(\mathcal{E})$ denotes the ICP with the same side information set $\mathcal{K}$ and demand vector $\mathbf{d}$, $B(\mathscr{I}(\mathcal{E}))$ represents the fitting matrix of the ICP $\mathscr{I}(\mathcal{E})$ as defined in \cite{BBJK1} and the notation $\hat{B} \sim B(\mathscr{I}(\mathcal{E}))$ is used to mean that $\hat{B}$ is a completion of $B(\mathscr{I}(\mathcal{E}))$.  For the ICP $\mathscr{I}(\mathcal{E})$ over $\mathbb{F}_q$, there are $q^{\sum_{i=1}^N|\mathcal{K}_i|}$ possible realizations $\hat{B}$ of its fitting matrix $ B(\mathscr{I}(\mathcal{E}))$. Corresponding to each of these realizations $\hat{B}$, a rank minimization is performed over the set of $q^{\sum_{i=1}^N|\mathcal{K}_i|^2}$ realizations of the side information matrix $A(\mathcal{E})$. For each realization $\hat{A}$ of $A(\mathcal{E})$, $(q^{\sum_{i=1}^N|\mathcal{K}_i|}+1)$ rank computations are made over $\mathbb{F}_q$ corresponding to finding the rank of $\hat{A}$ as well as the ranks of the  concatenated matrices $(\hat{A}|\hat{B})$ corresponding to the $q^{\sum_{i=1}^N|\mathcal{K}_i|}$ realizations $\hat{B}$ of $ B(\mathscr{I}(\mathcal{E}))$. Thus, for computing the minrank of an EICP  $\mathcal{E}(N,M,\mathcal{K},\mathbf{d})$ by the approach in \cite{MKR}, a total of  $q^{\sum_{i=1}^N|\mathcal{K}_i|^2} \times (q^{\sum_{i=1}^N|\mathcal{K}_i|}\ +1)$ rank computations are required. 
	
	Let us now look at the number of rank computations involved in determining the minrank of an EICP $\mathcal{E}(N,M,\mathcal{K},\mathbf{d})$ using the definition in this paper. In the set  $S = \{\mathbf{e_{d_i}} + \mathbf{v_i}\}_{i \in [N]}$, for each $i \in [N]$, the vector $\mathbf{v_i}$ which has support in $\mathcal{K}_i$, has $q^{|\mathcal{K}_i|}$ realizations. However to be included in the set over which rank minimization is performed, a particular realization of $(\mathbf{e_{d_i}} + \mathbf{v_i})$ should have support in the side information of some other user $u_j, \ j\neq i$. Hence out of the  $q^{\sum_{i=1}^N|\mathcal{K}_i|}$ possible realizations of the set $S$, the rank minimization is performed over a subset of them. Hence the number of rank computations is performed for computing the minrank is at most $q^{\sum_{i=1}^N|\mathcal{K}_i|}$. The computational complexity of searching whether there exists a user, such that the index set of its side information contains the support of a given realization of $(\mathbf{e_{d_i}} + \mathbf{v_i})$, is negligible compared to the complexity of rank computation and hence is ignored. This massive reduction in the number of rank computations required while going from the definition in \cite{MKR} to the definition in this paper is illustrated using the following numerical example. 
	\begin{example}
		\label{Ex:Minrank}
		Consider an EICP with $N =4$ users, $M = 4$ messages over $\mathbb{F}_2$, the side information at the users given by $\mathcal{K}_1 = \{2,3\}, \ \mathcal{K}_2 = \{1,3\}, \ \mathcal{K}_3 = \{4\}, \mathcal{K}_4 = \{1,2\}$ and the demand vector given by $\mathbf{d} = \{1,2,3,4\}$. Consider user $u_1$ which demands the message $x_1$. The contribution of $u_1$ to the set $\{\mathbf{e_{d_i}} + \mathbf{v_i}\}_{i \in [N]}$ are of the form $\mathbf{e_1} + \mathbf{v_1}$, where $\mathbf{v_1}$ can take values from the set $\{\mathbf{e_2}, \mathbf{e_3}, \mathbf{0}\}$ since $x_1 +x_2$ can be transmitted by $u_4$, $x_1 + x_3$ by $u_2$ and $x_1$ independently by either $u_2$ or $u_4$. Similarly $\mathbf{v_2}$ can take values from $\{\mathbf{e_1},\mathbf{e_3},\mathbf{0}\}$ whereas $v_3$ and $v_4$ can only take the value $\mathbf{0}$. Thus, the rank minimization is performed only over $9$ matrices including the $4 \times 4$ identity matrix and the eight matrices shown in Table \ref{Table:Minrank}. However, if we followed the approach in \cite{MKR}, it would be required to compute the ranks over $\mathbb{F}_2$ of $2^{\sum_{i=1}^N|\mathcal{K}_i|^2} = 2^{13} = 8192$ matrices of size $4 \times 7$ and $2^{\sum_{i=1}^N|\mathcal{K}_i|^2} \times 2^{\sum_{i=1}^N|\mathcal{K}_i|} =  2^{20} = 1048576$ matrices of size $4 \times 11$. 
		
		\begin{table*}[t]
			$	\begin{bmatrix}
				1 & 1 & 0 & 0 \\
				1 & 1 & 0 & 0 \\
				0 & 0 & 1 & 0 \\
				0 & 0 & 0 & 1
			\end{bmatrix}, \begin{bmatrix}
				1  & 0 & 0 & 0 \\ 
				1 & 1 & 0 & 0 \\ 
				0 & 1 & 1 & 0 \\ 
				0 & 0 & 0 & 1
			\end{bmatrix},\begin{bmatrix}
				1 & 0 & 0 & 0 \\ 
				1 & 1 & 0 & 0 \\ 
				0 & 0 & 1 & 0 \\ 
				0 & 0 & 0 & 1
			\end{bmatrix},\begin{bmatrix}
				1 & 1 & 0 & 0 \\ 
				0 & 1 & 0 & 0 \\ 
				1 & 0 & 1 & 0 \\ 
				0 & 0 & 0 & 1
			\end{bmatrix},\begin{bmatrix}
				1 & 0 & 0 & 0 \\ 
				0 & 1 & 0 & 0 \\ 
				1 & 1 & 1 & 0 \\ 
				0 & 0 & 0 & 1
			\end{bmatrix}, \begin{bmatrix}
				1 & 0 & 0 & 0 \\ 
				0 & 1 & 0 & 0 \\ 
				1 & 0 & 1 & 0 \\ 
				0 & 0 & 0 & 1
			\end{bmatrix}, \begin{bmatrix}
				1 & 1 & 0 & 0 \\ 
				0 & 1 & 0 & 0 \\ 
				0 & 0 & 1 & 0 \\ 
				0 & 0 & 0 & 1
			\end{bmatrix}, \begin{bmatrix}
				1 & 0 & 0 & 0 \\ 
				0 & 1 & 0 & 0 \\ 
				0 & 1 & 1 & 0 \\ 
				0 & 0 & 0 & 1
			\end{bmatrix}   $ 
			\caption{Set of possible matrices over which rank minimization is performed in Example \ref{Ex:Minrank}.}
			\label{Table:Minrank} 
		\end{table*}
		
	\end{example}
	
	%%%%%%%%%%%%%%%%%%%%%%%%%%%%%%%%%%%%%%%%%%%%%%%%%%%%%%%%%%%%%%%%%%%%%%%%%%%%%%%%%%%%%%%%%%%%%%%%%%%%%
	\section{A Graphical Representation of an EICP}
	\label{sec:Graph}
	
	In this section we propose a graphical representation of an EICP using directed bipartite graphs and explain why this representation is better than the graphical representation given by Potter and Wootters in \cite{PW}. The reader is assumed to be familiar with the graph theoretic terms and definitions given in the appendix.

	\subsection{Bipartite Problem Graph and Side Information Bipartite Graph}
	\label{subsec:PG}

	\begin{definition}[Bipartite Problem Graph]
		Let $\mathcal{E}(N,M,\mathcal{K},\mathbf{d})$ be an instance of the EICP. Its graphical representation called the bipartite problem graph is given by a directed bipartite graph $\mathcal{G}$ on the vertex set $V=(\mathcal{U},\mathcal{X})$ and the directed edge set $E = \{(u_i,x_j) : j \in \mathcal{K}_i\} \cup \{(x_i,u_j) : d_j = i \}$.
	\end{definition}
	The edges directed from the vertex set $\mathcal{U}$ to the vertex set $\mathcal{X}$ represent side information and the edges in the opposite direction represent the demanded messages. We use the notation $\mathcal{E}(\mathcal{G})$ to refer to the  EICP corresponding to a given bipartite  problem graph $\mathcal{G}$. 	
	
	For a system with $N$ users $\mathcal{U} = \{u_1,u_2, \cdots, u_N\}$, $M$ messages $\mathcal{X} = \{x_1,x_2,\cdots x_M\}$ and side information set $\mathcal{K}$ at the users, there could be $\prod\limits_{i=1}^{N}(M-|\mathcal{K}_i|)$ possible demand vectors and corresponding to each of these demands, there is an EICP. Let the set of this $\prod\limits_{i=1}^{N}(M-|\mathcal{K}_i|)$ arising from a side information set $\mathcal{K}$ be denoted as $\mathscr{E}_{\mathcal{K}}$.  The side information set $\mathcal{K}$ which is common to of all these EICPs  is represented using an undirected bipartite graph. 
	
	\begin{definition}[Side Information Bipartite Graph]
		A graphical representation of the side information set $\mathcal{K}$ is given by an undirected bipartite graph $\mathcal{G}_S$ on the vertex set $V=(\mathcal{U},\mathcal{X})$ and the  edge set $E = \{(u_i,x_j) : j \in \mathcal{K}_i\}$.
	\end{definition}
	
\begin{figure}[h] 		
		\centering
		\scalebox{0.8}{\includegraphics{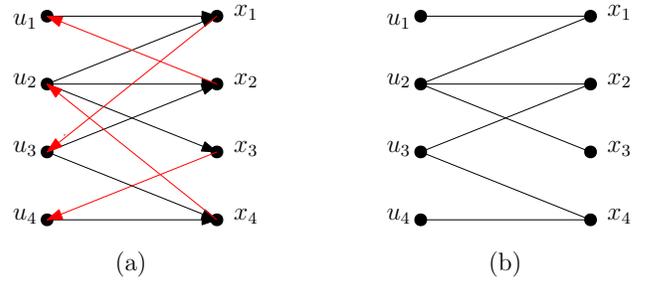}}
		\caption{Bipartite Problem  Graph and Side information Bipartite Graph of the EICP in Example \ref{ex:EICP1}}
		\label{fig:EICP1}
	\end{figure}

	Given a side information bipartite graph $\mathcal{G}_S$, the set of possible demand vectors is denoted as $D_{\mathcal{G}_S}$. Since a side information set $\mathcal{K}$ is analogous to the side information bipartite graph $\mathcal{G}_S$, the set of EICPs $\mathscr{E}_{\mathcal{K}}$ is also denoted as $\mathscr{E}_{\mathcal{G}_S}$. The EICP corresponding to a demand vector $\mathbf{d} \in D_{\mathcal{G}_S}$ is denoted as $\mathcal{E}(\mathcal{G}_S,\mathbf{d})$. The  bipartite problem graph of the EICP in Example \ref{ex:EICP1} is given in Fig. \ref{fig:EICP1}(a) and the corresponding side information bipartite graph is given in Fig. \ref{fig:EICP1}(b). 
	
	\begin{remark}
		A directed bipartite graph representation was given for an index coding problem in \cite{TDN}. For a single unicast ICP, if a set of $k$ nodes forms a clique in the side information graph, then their demands can be satisfied by a single transmission by the central server. The presence of structures like cycles or cliques can be seen from a side information graph. The hypergraph representation of a general ICP will reduce to a  side information graph for a single unicast ICP and hence patterns like cycles and cliques are not lost in this representation. Whereas, in the directed bipartite graph representation, due to the separation of the vertices into message vertices and receiver vertices, useful graph patterns like cliques will not be visible anymore. The bipartite representation does not give any extra information over the hypergraph representation, either.  
	\end{remark}
	
	\subsection{Comparison between the Problem Graph in \cite{PW} and the Bipartite Problem Graph}	
	\label{subsec:CompGraph}
	
	In \cite{PW}, an EICP was represented by a directed  graph $G$ which had a vertex set $V = \{v_{(u,x)}  : (u,x) \in \mathcal{U} \times \mathcal{X}$ and $x$ is demanded by $u \}$ and the edge set $E = \{(v_{(u,x)},v_{(w,y)}): u $ knows $y$ or $x = y\}$. It was further claimed that this representation is a generalization of the side information graph representation for a SUICP. While this is true and any general index coding problem can be represented using the graphical representation in \cite{PW}, there were already two other graphical representations of a general ICP reported in literature, the hypergraph representation introduced in \cite{DSC} and the bipartite representation in \cite{TDN}. There is no extra information that is conveyed by the representation of a general ICP in \cite{PW} over that conveyed by the other two  representations. Further, the problem graph in \cite{PW} does not differentiate between an ICP and a EICP. Also, the parameter derived from the problem graph and its equivalent problem matrix in \cite{PW} is the minrank for an ICP rather than any parameter of the EICP.

	The bipartite problem graph introduced here, on the other hand, is a more fitting representation for an EICP due to the following reasons. Unlike in an ICP, a coded transmission can involve only the messages in the side information of any one user. This information on which coded transmissions are possible is more clearly visible in the bipartite problem graph due to the separation of user and message vertices. The minrank of an EICP defined in \eqref{eq:minrank} can be derived from the bipartite problem graph $\mathcal{G}$ as the set of vectors $\{\mathbf{e_{d_i}} + \mathbf{v_i}\}_{i \in [N]}$ is same as a set $\{A_i\}_{i \in [N]}$ with  $A_i = N_{\mathcal{G}}^{-}(u_i) \cup S$, for a subset $S$ of the out-neighborhood of $u_i$, i.e., $S \subseteq N_{\mathcal{G}}^{+}(u_i)$. The condition that there should exist at least one user $u_j, j \neq i$ which could transmit the coded messages corresponding to each of the vectors in $\{\mathbf{e_{d_i}} + \mathbf{v_i}\}_{i \in [N]}$ can be translated to the condition that there should exists a user node $u_j \in \mathcal{U}$, $j \neq i$ such that the set $A_i$ is present in the out-neighborhood of $u_j$, i.e., $A_i \subseteq N_{\mathcal{G}}^{+}(u_j)$.
	
	The information that the same message is requested by multiple users conveyed by edges of the form $((u_i,x_j),(u_k,x_j))$ in the problem graph is present in the bipartite problem graph in the form of multiple outgoing edges from that particular message vertex. While there is no distinguishing factor between the side information edges and the edges corresponding to the same message being demanded by different users except the vertex labels in the problem graph in \cite{PW}, this information is clearly obtained in the direction of the side information and the demand edges in the bipartite problem graph. Further, the number of users demanding a given message has to be deciphered by looking at all the vertex labels and searching for a  common message in \cite{PW}, this information is easily conveyed by the out-degree of any message vertex. The splitting of a user node demanding multiple messages, say $k$ of them, into $k$ separate vertices happens in both the representations. Hence, no information is lost by going from the representation in \cite{PW} to our bipartite representation and there is a lot of additional clarity that is obtained from the bipartite representation as well. 
	
	Having seen the graphical representation of an EICP, we now identify certain graph structures which, if present in the side information bipartite graph, can result in savings in the number of transmissions required to solve the corresponding EICPs and a couple of transmission schemes utilizing these structures.

	\subsection{Graphical Structures in Single Unicast EICP}
	
	Similar to the definitions for ICP \cite{OngHo}, a single uniprior and single unicast EICPs are defined as follows.
	
	\begin{definition}[Single Uniprior EICP]
		An EICP $\mathcal{E}(N,M,\mathcal{K},\mathbf{d})$ is said to be single uniprior if
		\begin{enumerate}
			\item $|\mathcal{K}_i = 1|$, for all $i \in [N]$, and
			\item $\mathcal{K}_i \neq \mathcal{K}_j$, for $i \neq j$.
		\end{enumerate}
	\end{definition}
	
	Since in a single uniprior EICP, each user possesses only one message as side information, there is no opportunity for coding across messages and hence the only possible transmission scheme is to send all the requested messages independently. Thus, for a single uniprior EICP $\mathcal{E}$ with demand vector $\mathbf{d}$, the minrank $\kappa_q(\mathcal{E}) = uniq(\mathbf{d})$, where $uniq(\mathbf{d})$ is the number of distinct messages demanded in $\mathbf{d}$. In the rest of this section, we consider single unicast EICPs, which are defined as follows. 
	
	\begin{definition}[Single Unicast EICP]
		An EICP $\mathcal{E}(N,M,\mathcal{K},\mathbf{d})$ is said to be single unicast if 
		\begin{enumerate}
			\item $M =N$, and
			\item $d_i \neq d_j$, for $i \neq j$.
		\end{enumerate}
	\end{definition}

	Since in a single unicast EICP (SUEICP), the number of messages is equal to the number of users and each of the users demand a distinct message, without loss of generality, let us consider that in an SUEICP, user $u_i$ demands the message $x_i$, i.e., $d_i = i$. In the side information graph representation of a single unicast ICP, there are some graph structures like cycles and cliques which are useful in reducing the number of transmissions required. Transmission schemes based on these basic structures called the cycle cover scheme and its fractional version \cite{CASL} and clique cover scheme \cite{BiK} and its fractional version \cite{BKL} have been developed for index coding problems. Inspired by these covering schemes, here, we identify graph structures and covering schemes based on them for SUEICPs. 
	
	\begin{figure}[h] 		
		\centering
		\scalebox{0.8}{\includegraphics{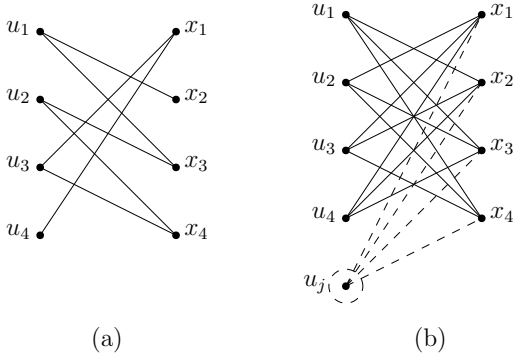}}
		\caption{Regular Tree and Bi-clique structures in SUEICP}
		\label{fig_SUEICP}
	\end{figure}
	
	Let an SUEICP with $N$ users and $N$ messages and the demand vector $\mathbf{d}$ such that $d_i = i$ be represented using its side information bipartite graph $\mathcal{G}_S$ and be denoted as $\mathcal{E}(N,\mathcal{G}_S)$.  Through the following definitions and lemmas, we give transmission schemes based on certain graph structures in $\mathcal{G}_S$.
	
	%  The definition of a class of SUICPs with symmetric consecutive and neighboring side information which was introduced in \cite{MCJ} and abbreviated as SNCS-SUICP is reproduced here for easy reference. 
	% 
	% \begin{definition}[SNCS-SUICP]
	% 	A single unicast ICP with $n$  users and $n$ messages such that the $i^{\text{th}}$ user $u_i$ demands the $i^{\text{th}}$ message $x_i$ is said to have symmetric neighboring consecutive side information if the side information at user $u_i$, $\mathcal{K}_i = \{x_{i+1},x_{i+2},\cdots,x_{i+D}\} \cup \{x_{i-1},x_{i-2},x_{i-U}\}$ such that $D+U < N$. 
	% \end{definition}
	% 
	% An SNCS-SUICP is said to be single-sided if $U = 0$. Consider an induced cycle on p vertices, $C_p \subset G$ on the user nodes as well as message nodes with indices $\{1,2,\cdots,p\}$. Here user $u_{i}$ requesting the message $x_{i}$ knows the message $x_{i+1}$. Hence, the induced cycle $C_p$ represents a one-sided SNCS-SUICP with $D=1$. The counterpart of a single sided SNCS-SUICP in embedded index coding was defined in \cite{MKR}. Since, in EICP, a user which has a single message cannot make a coded transmission, we consider the induced subgraph $C_{n,n}$ on $n$ user vertices and $n$ message vertices  of the bipartite problem graph of an SUEICP such that it represents an SNCS-EICP with $D=2$. 

	\begin{definition}[Regular Tree]
		\label{def:RTree}
		Consider a bipartite graph on the partite sets $A = \{a_1,a_2,\cdots,a_n\}$ and $B = \{b_1,b_2,\cdots,b_n\}$, $n \geq 3$, with the edge set, $E = \{(a_i, b_{i+1}), \ i \in [n]\} \cup \{(a_i, b_{i+2}), \ i \in [n-1]\}$, where $b_{n+1} = b_1$. Such a bipartite graph on $2n$ vertices, denoted as $T_{n,n}$, is called a regular tree. 
	\end{definition}
	
	\begin{remark}
		In the bipartite graph $T_{n,n}$, every node in $A$ except $a_n$ has degree two and similarly every node in $B$ except $b_2$ has degree two. The nodes $a_n$ and $b_2$ have degree one each. Thus, the bipartite graph $T_{n,n}$ on $2n$ vertices has a total of $2n-1$ edges and hence is a tree.  
	\end{remark}

	Fig. \ref{fig_SUEICP}(a) shows a regular tree $T_{4,4}$. Suppose $T_{4,4}$ represents the side information bipartite graph of an SUEICP with $N=M=4$ and the demand vector $\mathbf{d} = (1,2,3,4)$. If $u_1$ transmits $x_2+x_3$, $u_2$ transmits $x_3+x_4$, and $u_3$ transmits $x_4+x_1$, the demands of all the four users are met. Further, it can be verified that any set of two transmissions are not sufficient to satisfy the demands of all the $4$ users and hence the transmission scheme with $3$ transmissions is scalar linear optimal. We generalize this transmission scheme for an SUEICP on $4$ users to a scheme for an SUEICP on $N$ users in the following lemma. 
	
	\begin{lemma}
		\label{Lem:biCycle}
		If the side information bipartite graph $\mathcal{G}_S$ of an SUEICP with $N$ users and $N$ messages, $N \geq 3$, is a regular tree $T_{N,N}$, then $N-1$ transmissions are necessary and sufficient to satisfy the demands of all the users.
		\begin{proof}
			The set of user vertices and message vertices in $T_{N,N}$ is denoted by $\mathcal{U}_T = \{u_1,u_2,\cdots,u_N\}$ and $\mathcal{X}_T = \{x_1,x_2,\cdots,x_N\}$ respectively. From definition \ref{def:RTree}, we know that the first $N-1$ users in $\mathcal{U}_T$ knows two messages in $\mathcal{X}_T$ as side information and the side information at user $u_i$ , $i \in [N-1]$,  $\mathcal{K}_{i}$ is equal to $\{x_{i+1},x_{i+2}\}$ and that the side information at the user $u_N$ is $\mathcal{K}_N = \{x_{1}\}$ . Since we are considering an SUEICP, user $u_i$ demands the message $x_i$.  Consider the transmission scheme where each of the first $N-1$ users transmits the sum of the two messages in its side information, i.e., for $i \in [N-1]$,  user $u_i$ transmits $x_{i+1}+x_{i+2}$. 
			
			\emph{Decodability} : For every user $u_i \in \mathcal{U}_T\setminus\{u_1\}$ there exists a transmission of the form $x_{i} + x_{i+1}$, transmitted by the user $u_{i-1}$, from which it can decode $x_i$. Let the transmission made by user $u_i$ be denoted $T_i = x_{i+1} +x_{i+2}$. Consider the sum $T = T_1 -T_2 + T_3 - \cdots \pm T_{N-1}$ which is equal to $x_2 \pm x_1$ from which user $u_1$ can decode $x_1$. This proves that for an SUEICP on a regular tree $T_{N,N}$, $(N-1)$ transmissions are sufficient. Now we need to prove that $(N-1)$ transmissions are necessary.
			
			Suppose, we assume that $N-2$ transmissions are sufficient. Let the set of transmitting users be $\mathcal{U}_T \setminus \{u_j,u_N\}$, for some $j \in [N-1]$. So, the transmissions are $T_1 = x_2 +x_3$, $T_2 = x_3+x_4$, $\cdots, T_{j-1} = x_j + x_{j+1}$, $T_{j+1} = x_{j+2}+x_{j+3}, \cdots, T_{n-1}= x_N +x_1$. From these set of transmissions, it can be verified that user $u_{j+1}$ as well as $u_1$ cannot decode their requested messages. Since, this is true for any arbitrary user $u_j \in \mathcal{U}_T \setminus \{x_N\}$, the demands of all the $N$ users cannot be satisfied with $N-2$ transmissions or less. Hence, corresponding to an SUEICP with the side information bipartite graph being a regular tree $T_{N,N}$, $N-1$ transmissions are required
		\end{proof}	
	\end{lemma}

	\begin{definition}[Bi-clique]
		Consider a bipartite graph on the partite sets $A = \{a_1,a_2,\cdots,a_n\}$ and $B =\{b_1,b_2,\cdots,b_n\}$ where each of the node $a_i \in A$ is connected to all the nodes in $B\setminus\{b_i\}$ which implies that each node $b_i \in B$ is connected to all the nodes $A \setminus \{a_i\}$. This $n-1$ regular bipartite graph is called a ``bi-clique" and is denoted as $B_{n,n}$.
	\end{definition}  
	
	\begin{definition}[Covered Bi-clique]
		For a bi-clique $B_{n,n}$ on the partite sets $A$ and $B$, which is a subgraph of another bipartite graph $G$ on the partite sets $A_G \supset A$ and $B_G \supseteq B$, if there exists a node $a \in A_G \setminus A$ such that $a$ is connected to all nodes in  $B$, then the bi-clique is called a ``covered" bi-clique, denoted as $B_{n,n}^{c}$ and the node $a$ is called the covering node. 
	\end{definition}  
	
	Fig. \ref{fig_SUEICP}(b) shows a bi-clique $B_{4,4}$. Suppose it is a sub-graph on $4$ user nodes and $4$ message nodes of the side information bipartite graph $\mathcal{G}_S$ of an SUEICP. Assume that the user vertex $u_j$ and the edges coming from it shown in dotted lines are absent. Then, for the demand vector $\mathbf{d}$ such that for $i \in [4]$, user $u_i$ demands the message $x_i$, it can be seen that with two transmissions given by $T_1 = x_2+x_3+x_4$  transmitted by $u_1$ and $T_2 = x_1$ transmitted by $u_2$, the demands of all four users can be satisfied. However, if the user $u_j$ and the dotted edges incident on it are present in $\mathcal{G}_S$, then for the demand vector $\mathbf{d} =\{1,2,3,4\}$ corresponding to the users $\{u_1,u_2,u_3,u_4\}$, a single transmission $T_1 = x_1+x_2+x_3+x_4$ by user $u_j$ is sufficient to satisfy the demands of all the $4$ users.  In this scenario, the user $u_j$ is called the ``covering user".

	\begin{lemma}
		\label{Lem:clique}
		Consider a bi-clique $B_{n,n} \subseteq \mathcal{G}_S$ formed by the user nodes $\mathcal{U}_B = \{u_{i_1},u_{i_2},\cdots,u_{i_n}\}$ and message nodes $\mathcal{X}_B = \{x_{i_1},x_{i_2},\cdots,x_{i_n}\}$. For the SUEICP $\mathcal{E}(\mathcal{G}_S,\mathbf{d})$ corresponding to the demand vector $\mathbf{d}$ such that $d_i = i$, to satisfy the demands of the users in $\mathcal{U}_B$ we need $2- \mathcal{I}(B_{n,n}^{c})$ transmissions, where,  
		
		$\mathcal{I}(B_{n,n}^{c}) = \begin{cases}
			1 & \text{if } B_{n,n} \text{ is covered},  \\
			0 & \text{otherwise}
		\end{cases}$, is the indicator function on whether the bi-clique $B_{n,n}$ is covered or not.
		
		\begin{proof}
			The proof is straight forward. For a bi-clique $B_{n,n}$ on $n$ user nodes $\mathcal{U}_B = \{u_{i_1},u_{i_2},\cdots,u_{i_n}\}$ and $n$ message nodes $\mathcal{X}_B = \{x_{i_1},x_{i_2},\cdots,x_{i_n}\}$, a user $u_{i_j}$ knows all messages in $\mathcal{X}_B$ except the message it demands, i.e., $x_{i_j}$. Suppose there is a covering user $u_k$ for $B_{n,n}$, a single transmission, given by the sum of all messages in $\mathcal{X}_B$, transmitted by $u_k$ will satisfy the demands of all the users in $\mathcal{U}_B$. If there is no covering user, then $2$ transmissions $T_1 = \sum\limits_{x \in \mathcal{X}_B \setminus \{x_{i_1}\} }x$ by user $u_{i_1}$ and $T_2 = x_{i_1}$ by user $u_{i_2}$ will satisfy the demands of all the users in $\mathcal{U}_B$. Hence, the number of transmissions corresponding to a bi-clique is $1$ if a covering user is present and $2$ otherwise.
		\end{proof}
	\end{lemma}
	
	\begin{remark}
		A  cycle on $n \geq 3$ vertices in the side information graph of an SUICP is equivalent to a regular tree on $n$ user vertices and $n$ message vertices, $T_{n,n} \subseteq \mathcal{G}_S$ w.r.t number of transmissions required and similarly a clique on $n \geq 3$ vertices in the side information graph of an SUICP is equivalent to a covered bi-clique $B_{n,n}^c$ on $n$ user vertices and $n$ message vertices.
	\end{remark}
	
\begin{figure}[h]
		\begin{center}
			\includegraphics[scale=0.6]{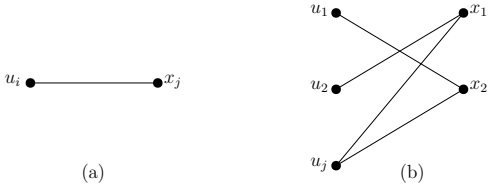}	
			\caption{Graph structures in $\mathcal{G}_S$ corresponding to both regular trees and covered bi-cliques for $n=1$ and $n=2$.}	
			\label{fig_BiCycle}	
		\end{center}	
	\end{figure}

	Having defined the graph structures and described how they provide savings in the required number of transmissions in an SUEICP, we now provide covering schemes based on these graph structures. Towards this end, we identify graph structures equivalent to regular trees and bi-cliques for $n=1$ and $n=2$. For $n=1$, both regular tree as well as a covered bi-clique is a single edge as shown in Fig. \ref{fig_BiCycle}(a) where the message is transmitted independently.  For $n=2$, a tree $T_{2,2}$, will not represent an SUEICP as for the graph $T_{2,2}$ to be connected, one of the two user nodes must know both the messages. The graph structure in an SUEICP which, similar to a cycle on two vertices in SUICP, requires one transmission to satisfy the demands of two users is a minimally connected graph on $3$ user nodes and $2$ message nodes as shown in Fig. \ref{fig_BiCycle}(b) which is also the covered bi-clique $B_{2,2}^c$. 
	
	\begin{lemma}[Tree Cover Scheme]
		\label{Lemma:BCCS}
		The tree-cover scheme for the SUEICP $\mathcal{E}(N, \mathcal{G}_S)$ identifies a maximal set of message-disjoint regular trees, say $\{T_{n_i,n_i}\}_{i=1}^K$, in $\mathcal{G}_S$ such that the union of these trees covers the message vertices of $\mathcal{G}_S$ . The total number of transmissions required to satisfy the demands of all the $N$ users in the SUEICP $\mathcal{E}(N, \mathcal{G}_S)$ using the tree cover scheme with $K$ regular trees, is $N-K+K_e$, where $K_e$ is the number of single edge trees.
		
		\begin{proof}
			Consider a maximal tree cover of the vertex set of $\mathcal{G}_S$ with $K$ regular trees with $0 \leq K_e \leq K$ of the trees being single edge trees corresponding to $n_i =1$. These require one transmission each, hence a total of $K_e$ transmissions. Corresponding to a regular tree $T_{n_i,n_i}$, $n_i >1$, there exists an embedded index code with number of transmissions equal to $n_i-1$ as explained in Lemma \ref{Lem:biCycle}. Since these trees partition the message set $\mathcal{X}$ of $\mathcal{G}_S$, $\sum\limits_{i=1}^K n_i = N$. The total number of transmissions made by users in all $K-N_e$ regular trees together is $\sum\limits_{i=1}^{K-K_e} (n_i-1) = N-K$. Hence, the total number of transmissions corresponding to the tree cover scheme is $N-K+K_e$.
		\end{proof}	
	\end{lemma}

	\begin{lemma}[Bi-Clique Cover Scheme]
		For an  SUEICP $\mathcal{E}(N, \mathcal{G}_S)$, a bi-clique cover scheme  identifies a minimal set of message-disjoint bi-cliques such that the union of these bi-cliques covers the message vertex set of $\mathcal{G}_S$. The total number of transmissions to solve the SUEICP $\mathcal{E}(N, \mathcal{G}_S)$ using the bi-clique cover scheme with $K$ message-disjoint bi-cliques $\{B_{n_i,n_i}\}_{i=1}^K$ is equal to $\sum\limits_{i=1}^K  (2- \mathcal{I}(B_{n_i,n_i}^{c}))$ which is bounded between $K$ and $2K$.
		\begin{proof}
			Corresponding to each bi-clique, $B_{n_i,n_i}$, there is a transmission scheme with either one or two transmissions depending the presence or absence of a covering user as explained in Lemma \ref{Lem:clique}. If a covering user is present for all the $K$ bi-cliques, only one transmission per bi-clique suffices giving a total of $K$ transmissions. On the other hand, if there is no covering user for any of the $K$ bi-cliques, corresponding to each of these bi-cliques, $2$ transmissions are required and hence giving a total of $2K$ transmissions. 
		\end{proof}
	\end{lemma}

	\begin{example}
		\label{ex:SUEICP}
		Consider the SUEICP with $M=N=7$ with side information as shown in Fig. \ref{fig_Covering} and where user $u_i$ demands the message $x_i$, for $i \in [M]$. For this SUEICP, the tree cover scheme identifies three regular trees as shown in Fig. \ref{fig_Covering}(a) with one transmission each given by $x_2 +x_3$ and $x_1+x_4$ for the first two regular trees and two transmissions $x_5+x_7$ and $x_6+x_7$ coresponding to the third regular tree thus requiring a total of $4$ transmissions to solve the SUEICP. For the same SUEICP, as shown in Fig. \ref{fig_Covering}(b), the bi-clique covering scheme identifies two bi-cliques, the first one covered and hence requiring a single transmission $x_1+x_2+x_3+x_4$ and the other requiring two transmissions $x_6+x_7$ and $x_5$, giving a total of $3$ transmissions to solve the SUEICP. It can be verified, the length of an optimal scalar linear solution for this SUEICP is $3$ which is achieved by the bi-clique covering scheme. 
		
		\begin{figure}[h]
			\begin{center}
				\includegraphics[scale=0.8]{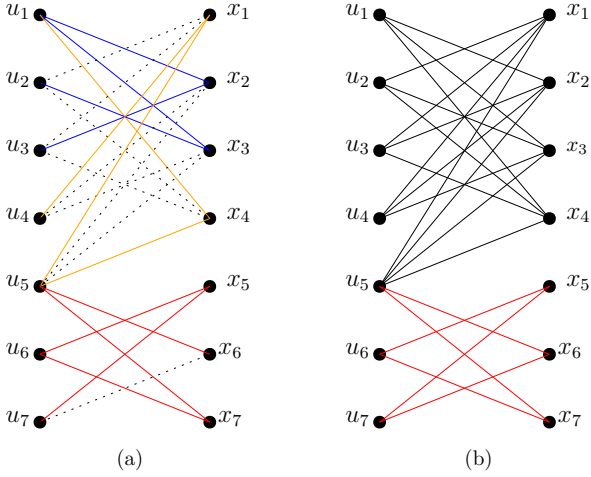}	
				\caption{Tree and Bi-clique Covering Schemes for the SUEICP in Example \ref{ex:SUEICP}}
				\label{fig_Covering}	
			\end{center}	
		\end{figure}
		
	\end{example}
	
	\begin{remark}
		If all the bi-cliques identified in the bi-clique covering scheme are covered bi-cliques, then it is the same as the covering scheme given by Algorithm 2 in \cite{PW}. A bi-clique which is not covered will be identified as two cliques in the problem graph by  Algorithm 2 in \cite{PW} and hence requires two transmissions same as that required by the bi-clique covering scheme in this paper.
	\end{remark}
	
	\begin{remark}
		While the solution obtained by the bi-clique covering scheme is a task-based solution as defined in \cite{PW}, the solution given by the tree covering scheme is not task-based, in general. 
	\end{remark}
	In the following section, we present a relation between the connectedness of the side information bipartite graph and minrank of an EICP in the following section. 
	%%%%%%%%%%%%%%%%%%%%%%%%%%%%%%%%%%%%%%%%%%%%%%%%%%%%%%%%%%%%%%%%%%%%%%%%%%%%%%%%%%%%%%%%%%%%%%%%%%%%%
	
	\section{Connectedness of the Bipartite Graph and Minrank}
	\label{sec:ConMinrank}
\begin{figure}[h] 		
		\centering
		\scalebox{0.8}{\includegraphics{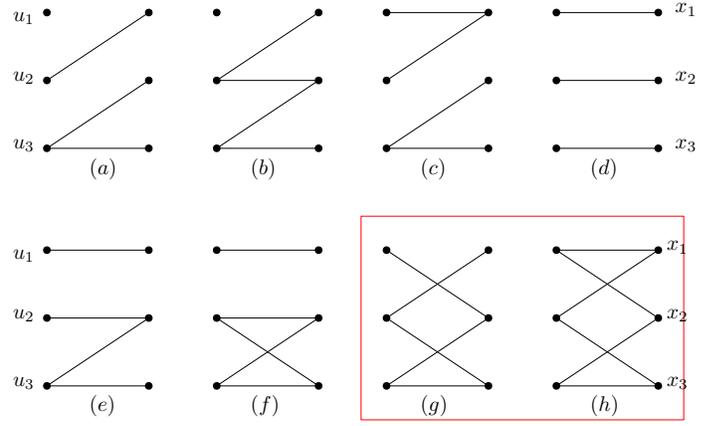}}
		\caption{Possible Side information Graphs  when $N = K = 3$}
		\label{fig:NK3}
	\end{figure}

	The motivation to look for a relation between the connectedness of the side information bipartite graph and the minrank of the corresponding EICPs  has been applications where the users have a lot of messages in common in their side information. Further, for the embedded index coding problems with $N=M=3$, the minrank is strictly less than the number of unique messages demanded only for problems with connected side information graphs and when all three messages are demanded. When $N = M = 3$, the set of all possible side information graphs upto user index and message index permutations are given in Fig. \ref{fig:NK3}. Among these patterns, only Fig. \ref{fig:NK3}(g) and \ref{fig:NK3}(h)  are connected. When all the three messages $x_1,x_2$ and $x_3$ are demanded, the number of transmissions required is less than three only for the EICPs with side information graphs as shown in either Fig. \ref{fig:NK3}(g) or \ref{fig:NK3}(h) up to index permutations. 
	
	Motivated by possible applications and the above observation regarding minrank and connectedness for the $N=M=3$ case, we state and prove a theorem connecting minrank $\kappa_q(\mathcal{G})$ and connectedness of $\mathcal{G}_S$. Towards that end, the following notations are defined. For a given  side information bipartite graph $\mathcal{G}_S$ with the partite sets $\mathcal{U}$ and $\mathcal{X}$, let $\mathcal{X}'$ denote the subset of $\mathcal{X}$ obtained by removing vertices of degree 1 in $\mathcal{X}$, i.e., $\mathcal{X}' = \mathcal{X} \setminus \{x_j : deg(x_j) = 1\}$ and the induced sub-graph on the vertex set $(\mathcal{U}, \mathcal{X}')$ be denoted by $\mathcal{G}'_S$. The set $\mathcal{X}'$ is the set of messages which could be possibly coded in the embedded index code as message nodes with degree 1 are present at only one user and cannot be coded and there are no degree zero message nodes as each message is assumed to be present at least at one user. For a demand vector $\mathbf{d}$, the unique messages demanded in $\mathbf{d}$  from the message set in $\mathcal{G}$ is denoted by $uniq(\mathbf{d}_{\mathcal{G}})$. 
	
	\begin{theorem}
		\label{Thm:ConMR}
		For a side-information bipartite graph $\mathcal{G}_S$ , for every demand vector  $\mathbf{d} \in D_{\mathcal{G}_S}$ such that $uniq(\mathbf{d}_{\mathcal{G}_S^{'}})  = |\mathcal{X}^{'}|$, the minrank of the EICP  $\mathcal{E}(\mathcal{G}_S, \mathbf{d})$ is strictly less than the number of distinct messages demanded, i.e., $\kappa_q(\mathcal{E}) < |uniq(\mathbf{d}_{\mathcal{G}_S})|$, if $\mathcal{G}_S$ is connected.  
		
		\begin{proof}
			
			If $\mathcal{G}_S$ is connected, $\mathcal{G}_S^{'}$ is also connected. For a given $\mathbf{d}$, it can be easily seen that $uniq(\mathbf{d}_{\mathcal{G}_S^{'}}) \leq uniq(\mathbf{d}_{\mathcal{G}_S})$. Let $|\mathcal{U}| = K$ and $|\mathcal{X}| = N$. The proof is by induction on the size of the set $\mathcal{X}^{'}$. 
			\begin{itemize}
				\item When $|\mathcal{X}^{'}| = 2$: Let $\mathcal{X}^{'} = \{x_1, x_2\}$.  Since $uniq(\mathbf{d}_{\mathcal{G}_S^{'}})  = |\mathcal{X}^{'}| = 2$, both $x_1$ and $x_2$ are demanded. Since $\mathcal{G}'_S$ is connected, every node in $\mathcal{U}$ is connected to at least one node in $\mathcal{X}^{'}$ and there exists at least one node in $\mathcal{U}$ which is connected to both$x_1$ and $x_2$. Hence, $x_1$ and $x_2$ can be coded together and transmitted in a single message as there exists at least one user which can send the coded message $x_1+x_2$ and every user demanding $x_1$ knows $x_2$ and vice-versa which implies that $\kappa_q(\mathcal{E}) = uniq(\mathbf{d}) - 1$.
				
				\item Induction Hypothesis : For $|\mathcal{X}^{'}| = k$, $\forall$ $\mathbf{d}$ s.t $uniq(\mathbf{d}_{\mathcal{G}_S^{'}})  = |\mathcal{X}^{'}| = k$, the theorem is assumed to be true.
				W.L.O.G, let the messages in $\mathcal{X}^{'}$ be $\{x_1,x_2,\cdots,x_k\}$. Let the set $D^{k}$ be defined as $ D^{k} \triangleq \{\mathbf{d} : uniq(\mathbf{d}_{\mathcal{G}_S^{'}})  = |\mathcal{X}^{'}| = k\}$ which is the set of demand vectors for which the theorem is true. 
				\item To prove : Theorem is true for $|\mathcal{X}^{'}| = k+1$. Let $\mathcal{X}^{'}$ be $\{x_1,\cdots,x_k,x_{k+1}\}$. and the set of all demand vectors for which we need to prove the theorem be denoted as $D^{k+1}$ $\triangleq \{\mathbf{d} : uniq(\mathbf{d}_{\mathcal{G}_S^{'}})  = |\mathcal{X}^{'}| = k+1\}$. The set of demand vectors in $D^k$ where the message $x_{k+1}$ is also demanded is defined as $D^{k}_{k+1}$ $\triangleq \{\mathbf{d} \in D^{k} \ s.t \ x_{k+1} \in \mathbf{d}\}$. Since by going from  $|\mathcal{X}^{'}| = k$ to $|\mathcal{X}^{'}| = k+1$, the degree of the message node $(x_{k+1})$ increased from $1$ to more than $1$, at least one more user now knows $x_{k+1}$. Hence, $D^{k+1} \subset D^{k}_{k+1} \subset D^{k}$. Since, the theorem is true for $D^k$ by induction hypothesis, it is true for the subset $D^{k+1}$. 
			\end{itemize}
		\end{proof}
	\end{theorem}
	\begin{corollary}
		For a connected side information graph $\mathcal{G}_S$, for all demand vectors where all the $M$ messages are demanded,  the number of transmissions required is strictly less than $M$, i.e.,  $ \forall \mathbf{d} \in D_{\mathcal{G}_S}$ such that $uniq(\mathbf{d}_{\mathcal{G}_S}) = M$, $\kappa_q(\mathcal{E}(\mathcal{G}_S,\mathbf{d})) < M$. 
	\end{corollary}
	
	\begin{remark}
		While Lemma \ref{Lem:biCycle} showed that for an SUEICP whose side information bipartite graph is a regular tree $T_{N,N}$, the number of transmissions required is $N-1$, by applying the corollary above, it can be seen that for any tree on $N$ user nodes and $N$ message nodes, the number of transmissions required for an SUEICP on these nodes is less than or equal to $(N-1)$.
	\end{remark}
	%\begin{theorem}
	%	\label{Thm:iff}
	%	For a side information graph $\mathcal{G}_U$ on  $N+M$ vertices with $N+M-1$ edges, for all 
	%\end{theorem}
	%%%%%%%%%%%%%%%%%%%%%%%%%%%%%%%%%%%%%%%%%%%%%%%%%%%%%%%%%%%%%%%%%%%%%%%%%%%%%%%%%%%%%%%%%%%%%%%%%%%%%

	%%%%%%%%%%%%%%%%%%%%%%%%%%%%%%%%%%%%%%%%%%%%%%%%%%%%%%%%%%%%%%%%%%%%%%%%%%%%%%%%%%%%%%%%%%%%%%%%%%%%%
	\section{Conclusion}
	\label{sec:Conc}
	This paper dealt with the problem of embedded index coding which is a distributed version of the well-studied index coding problem. An alternate definition of minrank of an EICP, which has a lesser computational complexity than the existing definition, and a proof that the minrank of an EICP characterized the length of an optimal scalar linear embedded index code for that EICP were given. A graphical representation of an EICP using bipartite graph, called the bipartite problem graph, was presented and shown to be a more fitting representation than the existing graphical representation as separating the users and messages into distinct vertices provided additional clarity. Further, from the bipartite representation, the possible combinations of messages which could be potentially coded together and transmitted could be easily identified besides being able to derive minrank of an EICP using it. For a single unicast EICP, graphical structures called regular trees and bi-cliques, in the side information bipartite graph, were identified  as counterpart structures of cycles and cliques in the  side information graph of a single unicast index coding problem and transmission schemes based on these structures were also presented. It was shown that connectedness of the side information bipartite graph helps in reducing the number of transmissions required to satisfy the demands of all the users in the corresponding EICPs.

While vector linear index coding has been extensively studied \cite{BKL,RSG}-\cite{MBVR} and shown to perform better, in general, than scalar solutions to ICPs \cite{RSG}, vector linear codes for EICPs haven't been looked at yet. It will be quite interesting to look at the vector linear solutions to EICPs and compare their performance with that of scalar solutions. Another problem of interest is to identify instances of EICP, especially those arising in practical scenarios, where we can characterize the minrank exactly and come up with transmission schemes with lengths matching the minrank as has been attempted in \cite{MKR,VICP}. Connectedness of the side information bipartite graph was shown to be a sufficient condition for reducing the minrank below the number of distinct messages demanded in Theorem \ref{Thm:ConMR}. However the requirement of connectedness is not necessary. It will be interesting and useful to derive necessary conditions relating connectedness of the bipartite side information graph to the minrank  of the corresponding EICPs. 
	
	%%%%%%%%%%%%%%%%%%%%%%%%%%%%%%%%%%%%%%%%%%%%%%%%%%%%%%%%%%%%%%%%%%%%%%%%%%%%%%%%%%%%%%%%%%%%%%%%%%%%%
	
	\section{Appendix}
	\textbf{Graph Theoretic Preliminaries} : 
	The following is a list of some basic graph theoretic definitions and notations  \cite{RD,DBW} that are used in this paper.
	A graph $G$ is a triple consisting of a vertex set $V(G)$, an edge set $E(G)$ and a relation that associates with each edge two vertices called its endpoints. 	Two vertices $u$ and $v$ are \emph{adjacent} or \emph{neighbors} in $G$ if there exist an edge $(u,v)$ in $G$. A \emph{clique} in a graph is a set of pairwise-adjacent vertices. 	An \emph{independent} set in a graph is a set of pairwise non-adjacent vertices. 	A graph $G$ is \emph{bipartite} if $V(G)$ is the union of two disjoint independent sets called partite sets of $G$. A graph $G'$ is called a sub-graph of the graph $G$, written as $G' \subseteq G$, if $V(G') \subseteq V(G)$ and $E(G') \subseteq E(G)$. If $G' \subseteq G$ and $G'$ contains all the edges $(x,y) \in E(G)$ with $x, y \in V(G')$, then G' is called an \emph{induced} sub graph of $G$. The set of neighbors of a vertex $v$ in $G$ is denoted as $N_G(v)$. More generally, for $U \subseteq V(G)$, the neighbors in $G$ of the vertices in $U$, is denoted by $N_G(U)$. The \emph{degree} of a vertex $v$, $deg(v)$, is the number of edges incident at it, which is equal to the number of neighbors of the vertex $v$ in the graph $G$. A vertex of degree $0$ is called an \emph{isolated} vertex. A \emph{path} is a non-empty graph $P = (V,E)$ of the form $V = \{x_0,x_1,\cdots,x_{k-1},x_k\}$ and $E = \{(x_0,x_1), (x_1,x_2),\cdots,(x_{k-1},x_k)\}$, where all the $x_i$s are all distinct. A \emph{cycle} is a path with the same first and last vertices being same. A graph $G$ is called \emph{connected} if it is non-empty and any two of its vertices are linked by a path in $G$. A maximal connected sub-graph of $G$ is a \emph{component} of $G$. The components are induced sub-graphs and their vertex sets partition $V(G)$. A \emph{minimally connected} graph $G$ (i.e., $G$ is connected and $G \setminus {e}$ is disconnected, for all edges $e \in G$) on $n$ vertices has $n-1$ edges. A minimally connected bipartite graph is called a tree.  A \emph{directed} graph is a graph with a direction associated with each edge in it. This implies that an edge $(u,v)$ is directed from the vertex $u$ to the vertex $v$ and an edge $(v,u)$ from $v$ to $u$ are two different edges in a directed graph whereas in an (undirected) graph, $(u,v)$ and $(v,u)$ mean the same edge between the endpoints $u$ and $v$. For a directed graph $G$, the out-neighborhood of a vertex $u$, denoted by $N_G^+(u)$, is set of vertices $\{v:(u,v) \in E(G)\}$. Similarly, the in-neighborhood of a vertex $u$, denoted by $N_G^-(u)$, is set of vertices $\{v:(v,u) \in E(G)\}$. A sub-graph $G'$ is said to \emph{cover} some other graph $G$ if it contains either all the vertices or all the edges  of the graph $G$. A \emph{Hamiltonian} cycle in a graph is a cycle which visits each of the vertices in the graph exactly once.
	%	\item A directed graph $G$ is said to be \emph{strongly connected} if for every pair of vertices $(u,v)$ in $G$, there exists a directed path from $u$ to $v$ as well as a directed path from $v$ to $u$. 
	
	%%%%%%%%%%%%%%%%%%%%%%%%%%%%%%%%%%%%%%%%%%%%%%%%%%%%%%%%%%%%%%%%%%%%%%%%%%%%%%%%%%%%%%%%%%%%%%%%%%%%%
	
	\section*{Acknowledgment}
	This work was supported partly by the Science and Engineering Research Board (SERB) of Department of Science and Technology (DST), Government of India, through J.C. Bose National Fellowship to Prof. B. Sundar Rajan.
	
	%%%%%%%%%%%%%%%%%%%%%%%%%%%%%%%%%%%%%%%%%%%%%%%%%%%%%%%%%%%%%%%%%%%%%%%%%%%%%%%%%%%%%%%%%%%%%%%%%%%%%%%%%%%%%
	
	%%%%%%%%%%%%%%%%%%%%%%%%%%%%%%%%%%%%%%%%%%%%%%%%%%%%%%%%%%%%%%%%%%%%%


\begin{thebibliography}{1}
		
		\bibitem{PW}
		A. Porter and M. Wootters, ``Embedded Index Coding," in \emph{IEEE Transactions on Information Theory}, vol. 67, no. 3, pp. 1461-1477, March 2021. 
		
		\bibitem{BiK}
		Y. Birk and T. Kol, "Informed-source coding-on-demand (ISCOD) over broadcast channels," Proceedings. IEEE INFOCOM '98, the Conference on Computer Communications. Seventeenth Annual Joint Conference of the IEEE Computer and Communications Societies. Gateway to the 21st Century (Cat. No.98, San Francisco, CA, USA, 1998, pp. 1257-1264 vol.3. 
		
		
		\bibitem{SMYA}
		S.~Li, M.~A.~Maddah-Ali, Q.~Yu and A.~S.~Avestimehr, ``A Fundamental Tradeoff Between Computation and Communication in Distributed Computing,'' in \textit{IEEE Transactions on Information Theory}, vol. 64, no. 1, pp. 109-128, Jan. 2018.
		
		\bibitem{JCM}
		M.~Ji, G.~Caire and A.~F.~Molisch, ``Fundamental Limits of Caching in Wireless D2D Networks,'' in \emph{IEEE Transactions on Information Theory}, vol. 62, no.2, pp. 849-869, Feb 2016.
		
		\bibitem{RSS}
		S. El Rouayheb, A. Sprintson and P. Sadeghi, "On coding for cooperative data exchange," 2010 IEEE Information Theory Workshop on Information Theory (ITW 2010, Cairo), Cairo, Egypt, 2010, pp. 1-5.
		
		\bibitem{BBJK1}
		Z. Bar-Yossef, Y. Birk, T. S. Jayram and T. Kol, "Index Coding with Side Information," 2006 47th Annual IEEE Symposium on Foundations of Computer Science (FOCS'06), Berkeley, CA, USA, 2006, pp. 197-206.
		
		\bibitem{BBJK}
		------------, ``Index Coding With Side Information,'' in \emph{IEEE Transactions on Information Theory}, vol. 57, no. 3, pp. 1479-1494, March 2011.
		
		\bibitem{DSC}
		S.~H. Dau, V.~Skachek, and Y.~M. Chee, ``Error correction for index coding with side information,'' in \emph{IEEE Transactions on Information Theory}, vol.~59, no.~3, pp. 1517--1531, Mar. 2013.
		
		
		\bibitem{TDN}
		A. S. Tehrani, A. G. Dimakis and M. J. Neely, ``Bipartite index coding," in \emph{2012 IEEE International Symposium on Information Theory Proceedings}, Cambridge, MA, USA, 2012, pp. 2246-2250.
		
		\bibitem{OngHo}
		L. Ong and C. K. Ho, ``Optimal index codes for a class of multicast networks with receiver side information," in \emph{Proc. 2012 IEEE International Conference on Communications (ICC)}, Ottawa, ON, Canada, 2012, pp. 2213-2218.
		
		\bibitem{RP}
		R. Peeters, “Orthogonal representations over finite fields and the chromatic number of graphs,” Combinatorica, vol. 16, no. 3, pp. 417–431, 	1996.
		
		\bibitem{MKR}
		A. A. Mahesh, N. Sageer Karat and B. S. Rajan, "Min-rank of Embedded Index Coding Problems," 2020 IEEE International Symposium on Information Theory (ISIT), Los Angeles, CA, USA, 2020, pp. 1723-1728.
		
		\bibitem{CASL}
		M. A. R. Chaudhry, Z. Asad, A. Sprintson, and M. Langberg, ``On the complementary index coding problem," in \emph{Proc.
			IEEE Int. Symp. Inf. Theory (ISIT)}, Jul. 2011, pp. 224–248.
		
		\bibitem{BKL}
		A. Blasiak, R. D. Kleinberg, and E. Lubetzky, ``Broadcasting with side information: Bounding and approximating the broadcast rate,'' in \emph{IEEE Trans. Inf. Theory}, vol. 59, no. 9, pp. 5811–5823, Sept. 2013.
		
		\bibitem{VANET}
		W. Zhu, D. Li and W. Saad, ``Multiple Vehicles Collaborative Data Download Protocol via Network Coding,” in \emph{IEEE Transactions on Vehicular Technology}, vol. 64, no. 4, pp. 1607-1619, April 2015. 
		
		\bibitem{VICP}
		Jesy P., N. S. Karat, Deepthi P.P., and B. S. Rajan, ``Index Coding in Vehicle to Vehicle Communication," in \emph{	IEEE Transactions on Vehicular Technology}, Vol.69, No. 10, Oct. 2020, pp.11926-11936.
		
%		\bibitem{TSG}
%		M. Tahmasbi, A. Shahrasbi and A. Gohari, ``Critical Graphs in Index Coding," in \emph{IEEE Journal on Selected Areas in Communications}, vol. 33, no. 2, pp. 225-235, Feb. 2015.
%		
%		\bibitem{ArK}
%		F. Arbabjolfaei and Y. Kim, "On critical index coding problems," 2015 IEEE Information Theory Workshop - Fall (ITW), Jeju, Korea (South), 2015, pp. 9-13.
%		
%		
		\bibitem{RSG}
		S. El Rouayheb, A. Sprintson and C. Georghiades, ``On the Index Coding Problem and Its Relation to Network Coding and Matroid Theory," in \emph{IEEE Transactions on Information Theory}, vol. 56, no. 7, pp. 3187-3195, July 2010
		
		\bibitem{MCJ}
		H. Maleki, V. R. Cadambe and S. A. Jafar, ``Index Coding—An Interference Alignment Perspective," in \emph{IEEE Transactions on Information Theory}, vol. 60, no. 9, pp. 5402-5432, Sept. 2014.
		
		\bibitem{MBVR}
		M. B. Vaddi and B. S. Rajan, ``Optimal vector linear index codes for some symmetric side information problems," 2016 IEEE International Symposium on Information Theory (ISIT), Barcelona, Spain, 2016, pp. 125-129.
		
		
		\bibitem{RD}
		R. Diestel, ``Graph Theory (Graduate Texts in Mathematics)", Springer, 2005.
		
		\bibitem{DBW}
		D. B. West, ``Introduction to Graph Theory", 2nd ed. Prentice Hall, 2000.
		
		
	\end{thebibliography}
\end{document}